\pdfoutput=1
\documentclass[journal]{IEEEtran}
\ifCLASSINFOpdf
\usepackage[pdftex]{graphicx}
\else
\usepackage[dvips]{graphicx}
\fi
\usepackage[cmex10]{amsmath}
\usepackage{array}
\usepackage{mdwmath}
\usepackage{mdwtab}
\usepackage{eqparbox}
\usepackage{cite}

\def\q{\quad}

\begin{document}

\title{Analytical expressions for the Electromagnetic Dyadic Green's Function in Graphene and thin layers}

\author{A.~Yu.~Nikitin, F.~J.~Garcia-Vidal, and~ L.~Martin-Moreno
\thanks{A.~Yu.~Nikitin and L.~Martin-Moreno are with the Instituto de Ciencia de Materiales de Arag\'{o}n and Departamento de F\'{i}sica de la Materia Condensada,
CSIC-Universidad de Zaragoza, E-50009, Zaragoza, Spain (e-mail: alexeynik@rambler.ru; lmm@unizar.es).}
\thanks{F.~J.~Garcia-Vidal is with the Departamento de F\'{i}sica Te\'{o}rica de la Materia Condensada, Universidad Aut\'{o}noma de Madrid, E-28049, Madrid,
Spain (e-mail: fj.garcia@uam.es).}
}

%

\maketitle

\begin{abstract}
An analytical general analysis of the electromagnetic Dyadic Green's Function for two-dimensional sheet (or a very thin film) is presented, with an emphasis on on the case of graphene.
A modified steepest descent treatment of the fields from a point dipole given in the form of Sommerfeld integrals is performed.
We sequentially derive the expressions for both out-of-plane and in-plane fields of both polarizations.
It is shown that the analytical approximation provided is very precise in a wide range of distances from a point source, down to a deep subwavelength region ($1/100$ of wavelength).
We separate the contribution from the pole, the branch point and discuss their interference. The asymptotic expressions for the fields are composed of the plasmon,
Norton wave and the components corresponding to free space.
\end{abstract}

\begin{IEEEkeywords}
Graphene, thin films, plasmon, Dyadic Green's Function.
\end{IEEEkeywords}

\IEEEpeerreviewmaketitle

\section{Introduction}

\IEEEPARstart{E}{lectromagnetic} properties of graphene have recently received a lot of attention due to a variety of application in photonics\cite{Bonaccorso10}. One of the attractive properties of graphene is its capacity to support highly-localized (nanometric) surface modes, i.e. graphene surface plasmons (GSPs) in terahertz (THz) and micro-wave frequency ranges
\cite{HansonAP08,HansonIEEE08,Engheta11,KoppensNL11,EfimovPRB11,StauberPRB11,NikitinPRB11,HansonAP11,PalomaPRB11,BasovNat12,KoppensNat12}. GSPs and their gate tunability open interesting possibilities for construction of tunable meta-materials \cite{Engheta11}, and merging photonics and electronics.  In particular, for applications related to strong light-matter interactions and biosensing, the interaction of a graphene sheet with a point emitter presents a special interest \cite{KoppensNL11,EfimovPRB11,StauberPRB11,NikitinPRB11,HansonAP11,PalomaPRB11,BasovNat12,KoppensNat12}. Localized excitation has allowed experimental demonstration of GSPs \cite{BasovNat12,KoppensNat12}. In comparison with other experimental techniques, local excitation of GSPs is more favorable due to very high values of GSP momentums.

The computation of patterns of the electromagnetic fields in graphene created by point source (as well as spontaneous emission rates), requires the knowledge of the Dyadic Green's Function (DGF) \cite{HansonAP08,HansonIEEE08,NikitinPRB11,HansonAP11,PalomaPRB11}. Its calculation involves notoriously difficult Sommerfeld-type integrals, with the integrands containing quickly oscillating functions, poles and branch cuts \cite{Novotny,FelsenMarcuvitz,Collin04}.

In this paper we will perform an analytical analysis of DGF, providing an asymptotic series expansion with the modified steepest-decent method \cite{FelsenMarcuvitz,Collin04}. We will show that this expansion, while being exact for long distances, in practice is very precise outside of its formal validity range. We will explicitly provide contributions for GSP, out-of-plane propagating waves and field components decaying algebraically along the graphene sheet. Notice that while all the examples correspond to the conductivity of graphene, the analytical expressions are applicable to any two-dimensional (2D) sheet.
The analytical expression can be useful for treatment of more complicated problems, related for instance to Lippmann-Schwinger integral equation \cite{Novotny}.

\begin{figure}[tbh!]
\includegraphics[width=8.3cm]{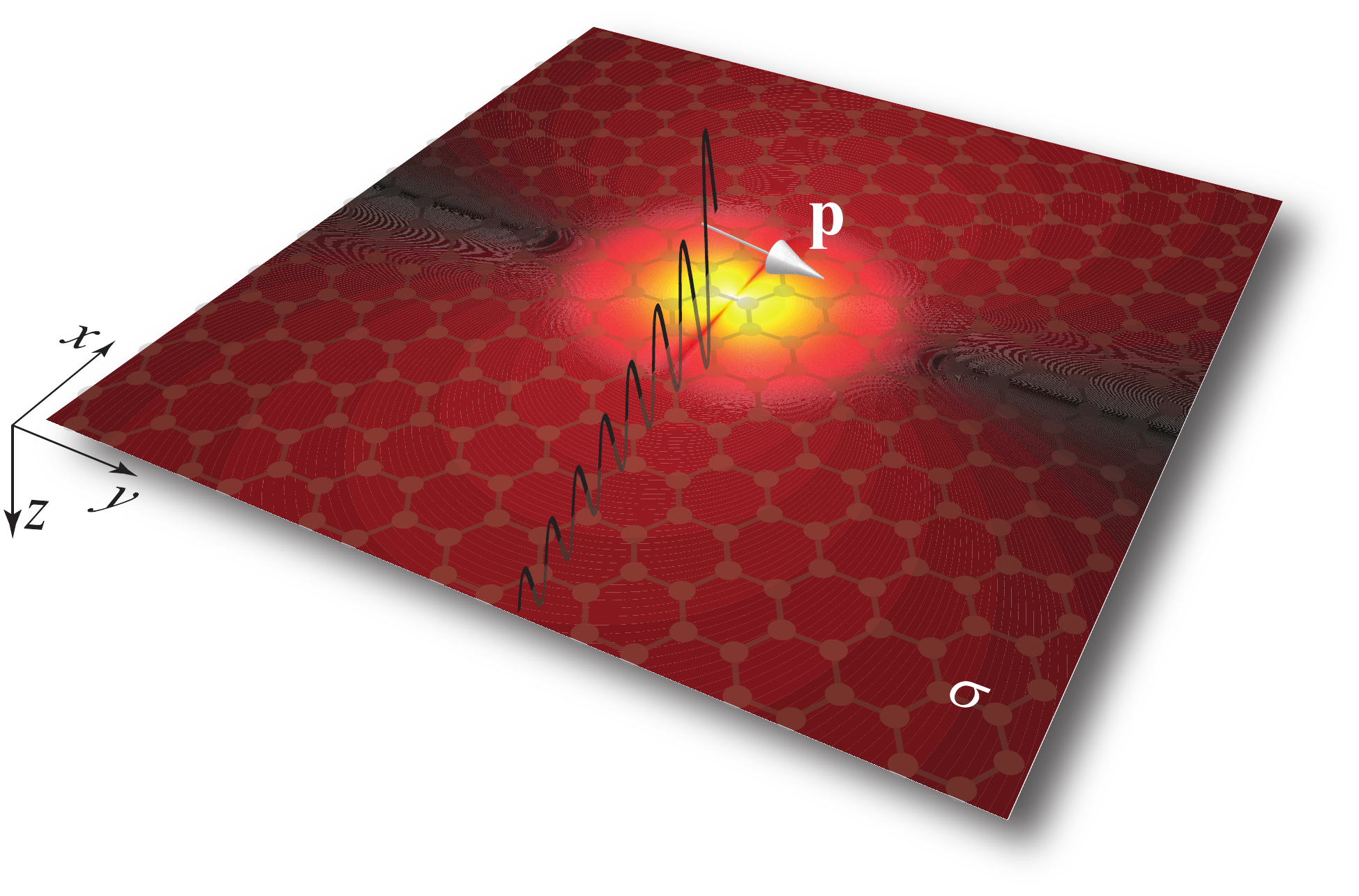}\\
\caption{The schematic of the studied system. A point source (dipole) is placed onto a graphene sheet, characterized by a two-dimensional conductivity $\sigma$. The colorplot represents an example of the spatial distribution of the electric field modulus for shown orientation of the dipole moment. Representative parameters for the graphene sheet corresponding to the formation of GSP are (see Appendix B for the definitions): $\Omega=0.4$, $T=300$ K, $\tau=1$ ps, $\mu=0.2$ eV.}\label{geom}
\end{figure}

\section{Formulation of the problem}

Let us consider a dipole with an arbitrary oriented dipole moment $\mathbf{p}$ placed at the point $(0,0,Z')$, with $Z'$ being the distance from a free-standing graphene sheet which covers the plane $Z=0$, see Fig.\ref{geom}. Without loss of generality, we will suppose that $Z'<0$. The time dependency is supposed to be $e^{-i\omega t}$, where $\omega$ is the angular frequency. Throughout the article, we express coordinates in the in-plane ($\mathbf{R}$) and normal  ($Z$) directions to the  graphene sheet in dimensionless units as $\mathbf{r}=k_\omega \mathbf{R}$ and $z=k_\omega Z$, with $k_\omega=2 \pi / \lambda$ being the free-space wavevector.

Graphene is represented by its in-plane complex conductivity $\sigma$. Throughout this paper, all examples are performed for $\sigma$ based on the random-phase-approximation \cite{Wunsch06,Hwang07,Falkovsky08}, see Appendix B.

The electric field $\mathbf{E}(\mathbf{r},z)$ emitted by our electric dipole, is given through the DGF $\hat{G}(\mathbf{r},z,z')=\hat{G}(\mathbf{r},z;\mathbf{r}'=0,z')$ by the following relation
\begin{equation}\label{f1}
\begin{split}
\mathbf{E}(\mathbf{r},z) = \hat{G}(\mathbf{r},z;z') \, \mathbf{p}(z').
\end{split}
\end{equation}
In the next section we will provide the exact general expressions for $\hat{G}(\mathbf{r},z;z')$.

\section{The general form of the Green's dyadic}

We have performed the analysis for a graphene placed onto the boundary of two different dielectrics. However, we have found that there is no much qualitative difference between this general case and free-standing (suspended) graphene. Since the analytical formula in the general case are quite lengthy, in this paper we will consider the DGF for a suspended graphene (taking also into account that free-standing samples are widely used in the experiments).

DGF satisfy the following differential equation
\begin{equation}\label{f2}
\begin{split}
\nabla \times\nabla \times \hat{G}(\mathbf{r},z;z') - k_\omega^2  \hat{G}(\mathbf{r},z;z') =
\hat{1}\delta(\mathbf{r})\delta(z-z') ,
  \end{split}
\end{equation}
where $\hat{1}$ is a diagonal unit matrix and $\delta$ is the Dirac delta function. The DGF must be complemented by the boundary conditions at the graphene sheet that we present below.

\subsection{Angular representation of DGF in Cartesian coordinates}

The solution for $\hat{G}$ can be can be expressed (see e.g. \cite{Novotny,FelsenMarcuvitz}) in terms of plane waves in free-space $\mathbf{u}_{\mathbf{q}\tau}e^{i\mathbf{qr}+iq_zz}$, characterized by their  in-plane momentum $\mathbf{q}=\mathbf{k}/k_\omega$ and polarization $\tau=\mathrm{TE, TM}$). The unitary vectors characterizing the polarization of each mode are:
\begin{equation}\label{f6}
\begin{split}
\mathbf{u}_{\mathbf{q}TE}^{\pm} = \frac{1}{q}
\begin{pmatrix}
-q_y\\
q_x\\
0
\end{pmatrix}, \q
\mathbf{u}_{\mathbf{q}TM}^{\pm} = \frac{q_z}{q}
\begin{pmatrix}
q_x\\
q_y\\
\mp\frac{q^2}{q_z}
\end{pmatrix}.
\end{split}
\end{equation}
where ``$+$'' (``$-$'') upward (downward) propagation along $z$ respectively.  $q_z=\sqrt{1-q^2}$ is the normalized z-component of the wavevectors.

The DGF reads
\begin{equation}\label{f3}
\begin{split}
\hat{G}(\mathbf{r},z;z')= \hat{G}_0(\mathbf{r},z;z') + \hat{G}_R(\mathbf{r},z;z'), \q z'<0, \q z<0,\\
\hat{G}(\mathbf{r},z;z')= \hat{G}_T(\mathbf{r},z;z'), \q z'<0, \q z>0,
\end{split}
\end{equation}
with $\hat{G}_0$ being the DGF in free space (FS),
\begin{equation}\label{f4}
\begin{split}
\hat{G}_0(\mathbf{r},z;z')=\sum_{\tau}\int \frac{d\mathbf{q}}{2q_z}\mathbf{u}^{\pm}_{\mathbf{q}\tau}\mathbf{u}_{\mathbf{q}\tau}^{\pm T}e^{i\mathbf{\mathbf{q}\mathbf{r}}+iq_z|z-z'|},
\end{split}
\end{equation}
and $\hat{G}_R$, $\hat{G}_T$  being the contributions due to the reflection and transmission in our 2D system
\begin{equation}\label{f5}
\begin{split}
&\hat{G}_R(\mathbf{r},z;z')=\sum_{\tau}\int \frac{d\mathbf{q}}{2q_z}R^\tau_{q}\mathbf{u}^{-}_{\mathbf{q}\tau}\mathbf{u}_{\mathbf{q}\tau}^{+ T}e^{i\mathbf{\mathbf{q}\mathbf{r}}-iq_z(z+z')},\\
&\hat{G}_T(\mathbf{r},z;z')=\sum_{\tau}\int \frac{d\mathbf{q}}{2q_z}T^\tau_{q}\mathbf{u}^{+}_{\mathbf{q}\tau}\mathbf{u}_{\mathbf{q}\tau}^{+ T}e^{i\mathbf{\mathbf{q}\mathbf{r}}+iq_z(z-z')}.
\end{split}
\end{equation}
In these expressions the superscript ``T'' stands for transposition.

In the above expressions $R^\tau_{q}$ and $T^\tau_{q}$ are the reflection and transmission coefficients for graphene.
These coefficients can be found by matching the magnetic $\mathbf{H}$ and electric $\mathbf{E}$ fields through the boundary conditions:
\begin{equation}\label{f7}
\begin{split}
&\mathbf{e}_z\times (\mathbf{E}_{-}-\mathbf{E}_{+}) = 0,\\
&\mathbf{e}_z\times(\mathbf{H}_{-}-\mathbf{H}_{+}) = \frac{4\pi}{c}\mathbf{j} = -\frac{4\pi}{c}\sigma\,\mathbf{e}_z\times[\mathbf{e}_z\times\mathbf{E_{+}}],
\end{split}
\end{equation}
where $\mathbf{E}_{-}$ ($\mathbf{H}_{-}$) and $\mathbf{E}_{+}$ ($\mathbf{H}_{+}$) stay for the electric (magnetic) fields in the regions of negative and positive $z$, respectively, and $\mathbf{e}_z$ is the unitary vector along the $+z$ direction. As a result of the matching we have
\begin{equation}\label{f8}
\begin{split}
&R^{TE}_{q} = \frac{-\alpha}{\alpha + q_{z}}, \q
R^{TM}_{q} = \frac{-\alpha q_z}{\alpha q_{z} + 1 },\\
&T_q^{TE} = \frac{q_{z}}{\alpha + q_{z}}, \q
T_q^{TM} = \frac{1 }{\alpha q_{z} + 1 },
\end{split}
\end{equation}
with $\alpha = 2\pi \sigma/c$, being the dimensionless 2D conductivity.

Explicitly, we have for $\hat{G}_0 = \hat{G}_0^{TM}+\hat{G}_0^{TE}$
\begin{equation}\label{f9}
\begin{split}
\hat{G}_0^{TE}(\mathbf{r}) =\frac{ik_\omega}{8\pi^2}
\int\dfrac{d\mathbf{q}}{q_{z}q^2}e^{i\mathbf{q}\mathbf{r}+iq_z|z-z'|}\\
\times\begin{pmatrix}
q_y^2 & -q_xq_y & 0\\
-q_xq_y & q_x^2 & 0\\
0 & 0 & 0\\
\end{pmatrix},\\
\hat{G}_0^{TM}(\mathbf{r}) =\frac{ik_\omega}{8\pi^2}
\int\dfrac{d\mathbf{q}}{q^2}e^{i\mathbf{q}\mathbf{r}+iq_z|z-z'|}\\
\times\begin{pmatrix}
q_x^2q_{z} & q_xq_yq_{z} & \mp q_xq^2\\
q_xq_yq_{z} & q_y^2q_{z} & \mp q_yq^2\\
\mp q_xq^2 & \mp q_yq^2 & q^4/q_{z}\\
\end{pmatrix}.
\end{split}
\end{equation}
Analogously, $\hat{G}_R = \hat{G}_R^{TM}+\hat{G}_R^{TE}$  reads
\begin{equation}\label{f10}
\begin{split}
\hat{G}_R^{TE}(\mathbf{r}) =\frac{ik_\omega}{8\pi^2}
\int\dfrac{d\mathbf{q}}{q_{z}q^2}R^{TE}_{q}e^{i\mathbf{q}\mathbf{r}-iq_z(z'+z)}\\
\times\begin{pmatrix}
q_y^2 & -q_xq_y & 0\\
-q_xq_y & q_x^2 & 0\\
0 & 0 & 0\\
\end{pmatrix},\\
\hat{G}_R^{TM}(\mathbf{r}) =\frac{ik_\omega}{8\pi^2}
\int\dfrac{d\mathbf{q}}{q^2}R^{TM}_{q}e^{i\mathbf{q}\mathbf{r}-iq_z(z'+z)}\\
\times\begin{pmatrix}
q_x^2q_{z} & q_xq_yq_{z} & -q_xq^2\\
q_xq_yq_{z} & q_y^2q_{z} & -q_yq^2\\
q_xq^2 & q_yq^2 & -q^4/q_{z}\\
\end{pmatrix},
\end{split}
\end{equation}
and finally, for the transmission part $\hat{G}_T = \hat{G}_T^{TM}+\hat{G}_T^{TE}$ we have
\begin{equation}\label{f11}
\begin{split}
\hat{G}_T^{TE}(\mathbf{r}) =\frac{ik_\omega}{8\pi^2}
\int\dfrac{d\mathbf{q}}{q_{z}q^2}T^{TE}_{q}e^{i\mathbf{q}\mathbf{r}+iq_z(z'+z)}\\
\times\begin{pmatrix}
q_y^2 & -q_xq_y & 0\\
-q_xq_y & q_x^2 & 0\\
0 & 0 & 0\\
\end{pmatrix},\\
\hat{G}_T^{TM}(\mathbf{r}) =\frac{ik_\omega}{8\pi^2}
\int\dfrac{d\mathbf{q}}{q^2}T^{TM}_{q}e^{i\mathbf{q}\mathbf{r}+iq_z(z'+z)}\\
\times\begin{pmatrix}
q_x^2q_{z} & q_xq_yq_{z} & -q_xq^2\\
q_xq_yq_{z} & q_y^2q_{z} & -q_yq^2\\
-q_xq^2 & -q_yq^2 & q^4/q_{z}\\
\end{pmatrix}.
\end{split}
\end{equation}

Eqs. \eqref{f9}-\eqref{f11} present the angular representation of DGF in Cartesian coordinates. By transforming a cylindrical coordinate system, these expression
can be greatly simplified.

\begin{figure}[tbh!]
\includegraphics[width=6cm]{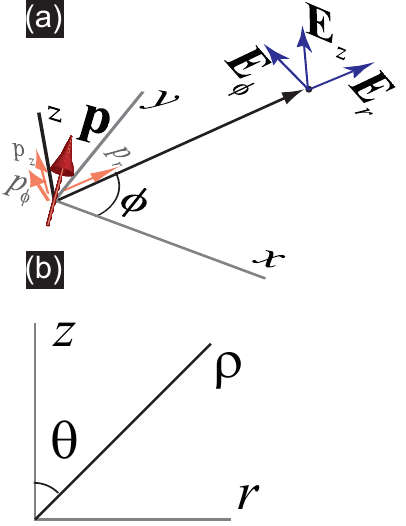}\\
\caption{Transformation of the coordinates.}\label{transform}
\end{figure}

\subsection{DGF in cylindrical coordinates}

As was previously shown, the Purcell factor (the total decay rate normalized to the free space decay rate) of the point emitter placed directly over the graphene monolayer diverges due to losses through the evanescent waves with large $q$-components \cite{EfimovPRB11,StauberPRB11,PalomaPRB11}. This means that the dipole placed directly on the graphene surface is quenched. However, the parameter that accounts for the efficiency of the
coupling to GSP, ($\beta$-factor, defined as the ratio of
the emitter's decay rate through GSP to its total decay rate) has an optimum value in the region of very small distances from the dipole to graphene: $|z'_{opt}|/\lambda\sim 10^{-2}$ (see \cite{PalomaPRB11}). Then, taking into account that the problems related to the high values of $\beta$-factor are relevant, the range of small distances presents a special interest. Another important point is that the field patterns at the distances $r>|z'|$ do not differ essentially upon the field patterns created by a dipole lying directly on the monolayer. Therefore, for the above two reasons, in this paper we will consider the dipole placed directly onto the graphene sheet.

We would like to notice, that some physical systems can be reduced to a problem of a dipole lying directly on the graphene sheet. For instance, a subwavelength aperture in graphene sheet can be represented by an effective dipole placed directly onto the sheet, (for comparison with the case of metal films see \cite{NikitinPRL10,JohanssonPRB11}).

Additionally, since the analytical treatment of both reflection ($z<0$) and transmission ($z>0$) parts of the DGF is similar, we will derive the expressions for the transmission part, $\hat{G}_T$.

So, supposing that $z'=0^-$ in the previous expressions, the Green's dyadic $\hat{G}(\mathbf{r},z)\equiv\hat{G}_T(\mathbf{r},z,z'=0^-)$ can be simplified.
The symmetry of the problem makes it convenient to work in cylindrical coordinates $(r,\phi,z)$, see Fig.~\ref{transform}~(a):
\begin{equation}\label{f12}
\begin{split}
x = r \cos\phi, \q y = r \sin\phi, \q z=z.
  \end{split}
\end{equation}
In this system of coordinates the Green's dyadic can be obtained from the one in cartesian coordinates through
\begin{equation}\label{f13}
\begin{split}
\hat{G}^{cyl} = \hat{T}^{-1}\,\hat{G}^{cart}\,\hat{T},
\end{split}
\end{equation}
where
\begin{equation}\label{f14}
\begin{split}
\hat{T} =
\begin{pmatrix}
\cos\phi & -\sin\phi & 0\\
\sin\phi & \cos\phi & 0\\
0 & 0 & 1\\
\end{pmatrix}.
  \end{split}
\end{equation}
Having performed this transformation, the DGF in cylindrical coordinates is expressed in the form of Sommerfeld integrals as $\hat{G} = \hat{G}_p+\hat{G}_s$:
\begin{equation}\label{f15}
\begin{split}
\hat{G}_s(r,z)= \frac{ik_\omega}{8\pi} \int_0^\infty \, \frac{dqq}{\alpha+q_{z}}  \,
e^{iq_zz}\\
\times\begin{pmatrix}
J_+(qr)& 0 & 0\\
0 & J_-(qr) & 0\\
0 & 0 & 0\\
\end{pmatrix}\\
\hat{G}_p(r,z)= \frac{ik_\omega}{8\pi} \int_0^\infty \, \frac{dq}{\alpha q_{z} + 1 }  \,
 e^{iq_zz}\\
\times\begin{pmatrix}
qq_{z}J_-(qr) & 0 & -2 i q^2J_1(qr)\\
0 & qq_{z}J_+(qr) & 0\\
-2 i q^2J_1(qr) & 0 & \frac{2 q^{3}}{q_{z}} J_0(qr)\\
\end{pmatrix},
\end{split}
\end{equation}
where the subscripts ``s'' and ``p'' correspond to TE and TM polarizations respectively. In this expressions, $J_\pm(qr)= J_0(qr)\pm J_2(qr)$ and $J_n(qr)$ are Bessel functions of $n$th order.  Equations \eqref{f15} can be treated numerically. We show some recipes for the integration in the complex plane $q$ in Appendix A. Let us now proceed with the asymptotic expansion of the DGF.

\section{Asymptotic expansion of the DGF}

In this section we will derive explicit asymptotic expressions for the elements of DGF following the steepest-decent method, modified in order to take into account the presence of both poles and branch points close to the integration path \cite{FelsenMarcuvitz,NikitinPRL10}.

First, using the identity
\begin{equation}\label{a1}
2J_n = H^{(1)}_n + H^{(2)}_n,
\end{equation}
where $H_n$ are Hankel functions, we can extend the limit of integration to the whole real $q$-axis in \eqref{f15}
\begin{equation}\label{a2}
\int_0^\infty \, dq\, J_n(qr)F(q)= \frac{1}{2}\int_{-\infty}^\infty \, dq\, H^{(1)}_n(qr)F(q),
\end{equation}
where according to Eqs.~\eqref{f15} function $F(q)$ is odd/even for even/odd values of $n$.
In \eqref{a2} we have used the symmetry of the Hankel functions $H^{(2)}_n\left(-x\right) = -e^{in\pi}H^{(1)}_n(x)$.

Second, we use the asymptotic form of the Hankel functions for large arguments ($qr\gg1$). Notice that the region that provides the major contribution to the integral corresponds to $q\geq1$. Then the formal condition of the asymptotic expansion validity for the lower value of the contributing $q$ reads as $r\gg1$. However, as we will show below, the true region of distances where the asymptotic approximation is valid is much less restricted.
Retaining the first two terms in $H^{(1)}_n(qr)$, the expansion reads
\begin{equation}\label{a3}
\begin{split}
H^{(1)}_n(x) =\sqrt{\frac{2}{\pi x}}e^{i\left[x-\frac{\pi}{2}(n+\frac{1}{2})\right]}\left(1 + i\frac{4n^2-1}{8x}\right) + O(x^{-\frac{5}{2}}).
\end{split}
\end{equation}
For convenience, let us normalize the DGF as follows
\begin{equation}\label{a4}
\begin{split}
\hat{G} = \frac{k_\omega e^{i\frac{\pi}{4}}}{8\pi}\sqrt{\frac{2}{\pi r}}\,\hat{g}.
\end{split}
\end{equation}
Then using \eqref{a1}-\eqref{a2}, we find from \eqref{f15} the expressions for $\hat{g}=\hat{g}_s+\hat{g}_p$:
\begin{equation}\label{a5}
\begin{split}
\hat{g}_\tau(r,z) = \int\limits_{-\infty}^\infty dq \,\frac{e^{iqr+iq_zz}}{f_{\tau}(q)} \left[\hat{A}_\tau^{(1)}(q) + \frac{i}{r}\hat{A}_\tau^{(2)}(q)\right].
\end{split}
\end{equation}
with $\tau=s,p$ and the superscripts $(1,2)$ of $A$ indicate distinct dependencies upon $r$.
The denominators in the integral are
\begin{equation}\label{a5.2}
\begin{split}
f_{s}(q) = \alpha + q_{z}, \,\, f_{p}(q) = \alpha q_{z} + 1,
\end{split}
\end{equation}
and nominators
\begin{equation}\label{a6}
\begin{split}
&\hat{A}_p^{(1)}(q) = \sqrt{q}\left[q_z\hat{r}\hat{r}-q\left(\hat{r}\hat{z}+\hat{z}\hat{r}\right)+\frac{q^2}{q_z}\hat{z}\hat{z}\right],\\
&\hat{A}_s^{(1)}(q) = \sqrt{q}\hat{\phi}\hat{\phi},\\
&\hat{A}^{(2)}_p = \frac{1}{8\sqrt{q}}\left[7q_z\hat{r}\hat{r}- 8q_z\hat{\phi}\hat{\phi}-3q\left(\hat{r}\hat{z}+\hat{z}\hat{r}\right)-\frac{q^2}{q_z}\hat{z}\hat{z}\right],\\
&\hat{A}^{(2)}_s = \frac{1}{\sqrt{q}}\left(-\hat{r}\hat{r}+\frac{7}{8}\hat{\phi}\hat{\phi}\right).
\end{split}
\end{equation}
where $\hat{r}$, $\hat{\phi}$, $\hat{z}$ represent the unit vectors.

\begin{figure}[tbh!]
\includegraphics[width=8.3cm]{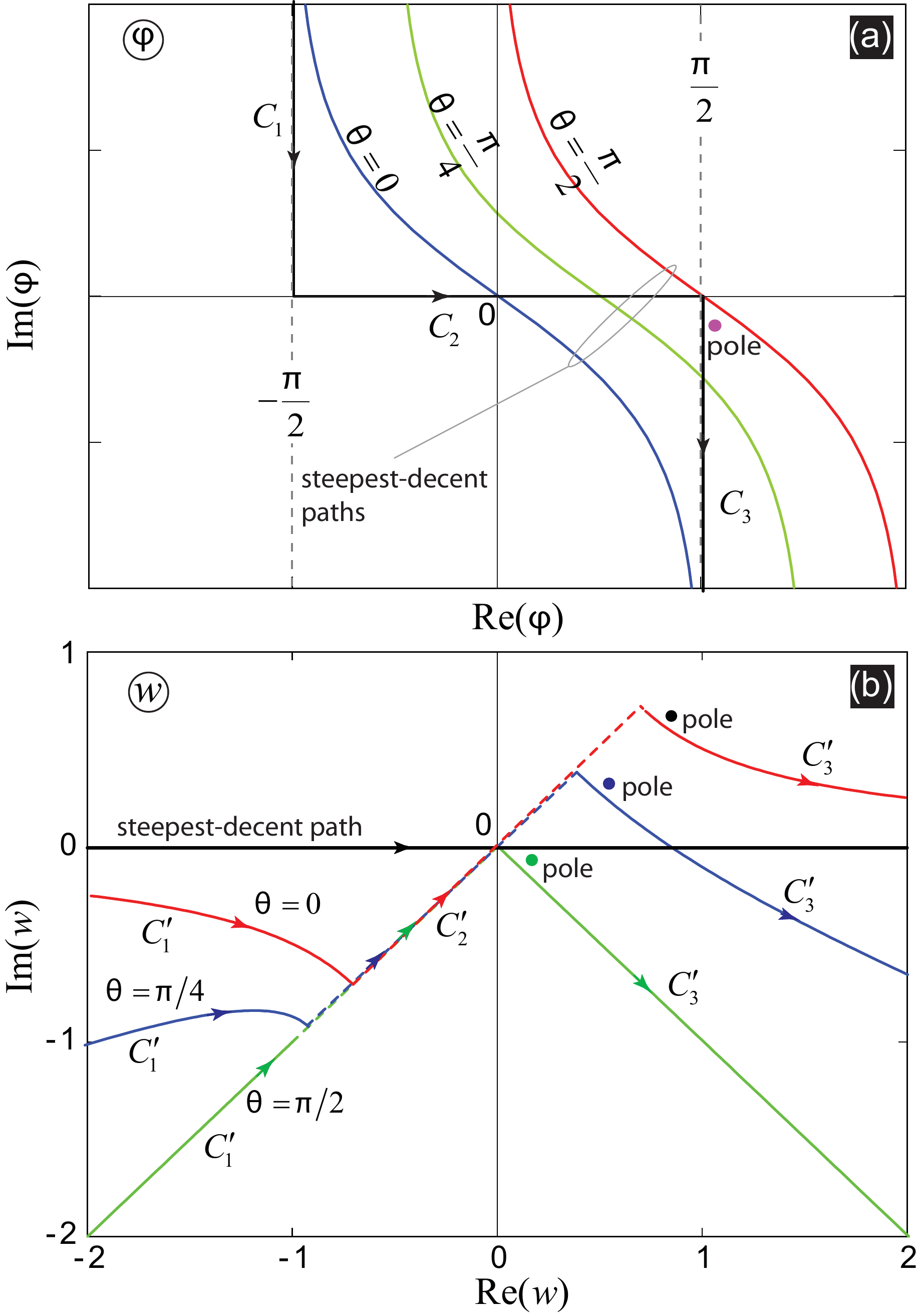}\\
\caption{Integration contours in the complex plane of the integration variables $\varphi$ [panel (a)] and $w$ [panel (b)] for different values of angle $\theta$. In panels (a) and (b) the initial integration contour corresponds to $C=C_1+C_2+C_3$ and $C'=C'_1+C'_2+C'_3$ respectively. The pole position is shown by a circular symbol. In the complex plane of $\varphi$ the initial path is the same and the steepest-decent path depends upon $\theta$, while in the complex plane $w$ the steepest-decent path is the same, $\mathrm{Im}(w)=0$, and the initial integration path changes with $\theta$. The position of the pole in the complex plane $w$ is dependent upon $\theta$.}\label{integr}
\end{figure}

As we see, the integrands contain branch points corresponding to $q_{z}$, and branch cuts corresponding to $\mathrm{Im}(q_z)=0$.
In order to remove these problematic peculiarities, we perform the following standard change of integration variable and coordinates (see Fig.~\ref{transform}~(b)):
\begin{equation}\label{a7}
q = \sin\varphi, \q q_z = \cos\varphi, \q r = \rho\sin\theta, \q z = \rho\cos\theta.
\end{equation}
The integrals \eqref{a5} then transforms as
\begin{equation}\label{a7.1}
\begin{split}
&\int\limits_{-\infty}^\infty dq \,e^{iqr+iq_zz}\frac{\hat{A}^{(n)}_\tau(q)}{f_\tau(q)} \\
&=\int_C d\varphi e^{i\rho\cos(\varphi-\theta)}\cos\varphi\frac{\hat{A}^{(n)}_\tau(\sin\varphi)}{f_\tau(\sin\varphi)} ,
\end{split}
\end{equation}
where the integration contour $C$ pases through the complex plane $\varphi$, see Fig.~\ref{integr}~(a) and corresponds to the real axis in the complex $q$-plane. At this stage we can perform a steepest-decent integration. Then the integration path must be transformed to $\cos[\mathrm{Re}(\varphi)-\theta]\cosh[\mathrm{Im}(\varphi)] = 1$, see Fig.~\ref{integr}~(a), with the saddle point $\varphi=\theta$. However, the steepest-decent integration is much easier in another complex variable plane. This variable  $w$ is given as follows
\begin{equation}\label{a7.2}
w = \sqrt{2}e^{i\frac{\pi}{4}}\sin\left(\frac{\varphi-\theta}{2}\right),
\end{equation}
so that $i\cos(\varphi-\theta) = i - w^2$. With this change of variable $\hat{g}_\tau$ becomes
\begin{equation}\label{a8}
\begin{split}
&\hat{g}_\tau = e^{i \rho}\int_{C'} dw e^{- \rho w^2}\left[\hat{\Phi}_\tau^{(1)}(w)+\frac{i}{r}\hat{\Phi}_\tau^{(2)}(w)\right], \\
&\hat{\Phi}_\tau^{(n)}(w) = \cos[\varphi(w)]\cdot\frac{d\varphi}{dw}\cdot\frac{\hat{A}_\tau^{(n)}[q(w)]}{f_\tau[q(w)]},
\end{split}
\end{equation}
where the integration contour $C'$ is shown in see Fig.~\ref{integr}~(b). The steepest-decent integration path is now simply given by $\mathrm{Im}(w)=0$, and the saddle point is located in the origin, $w=0$. Notice that the branch points $q=\pm1$ (corresponding to $q_z=0$)  are given by $\varphi=\pm\pi/2$ in the complex plane $\varphi$, while in the $w$-plane they are located at $w = \sqrt{2}e^{i\frac{\pi}{4}}\sin\left(\pm\frac{\pi}{4}-\frac{\theta}{2}\right)$. For $\theta=\pm\pi/2$ the branch point and saddle point in $w$-plane coincide.

An important point here is that the elements of the integrant dyadic in $\hat{g}_\tau$ are singular due to the presence of the poles  [$f_{p}(q)=0$ and  $f_{s}(q)=0$] in the denominators. These poles are located at $q=q_p$ and  $q=q_s$ respectively. They correspond to TM-surface wave (GSP) and TE surface wave \cite{HansonAP08}
\begin{equation}\label{a9}
\begin{split}
q_p =\sqrt{1-\frac{1}{\alpha^2}}, \\
q_s =\sqrt{1-\alpha^2}.
\end{split}
\end{equation}
The positions of the poles depend essentially upon the value of the normalized conductivity $\alpha$. TM waves correspond to $\mathrm{Im}(\alpha)>0$, while TE ones correspond to  $\mathrm{Im}(\alpha)<0$, so that TM and TE surface waves cannot exist at the same frequency see \cite{HansonAP08}. High values of $|\alpha|$ correspond to large $q_s$ while low values of $|\alpha|$ correspond to large $q_p$. In principle, taking into account a wide range of metamaterials that are available at present, a wide range of $\alpha$ is also accessible. A mathematical treatment of the problem corresponding to a three-dimensional (3D) layer of a very thin thickness $h\ll\lambda$ with the dielectric permittivity $\varepsilon_{3D}$ can be reduced to the case of a 2D sheet with an effective 2D normalized conductivity $\alpha_{\mathrm{eff}}$. The relation between 2D effective conductivity and 3D permittivity is established from the comparison of the Fresnel coefficients and reads as $\alpha_{\mathrm{eff}} = \pi h\, \varepsilon_{3D}/i\lambda$.

Returning to the case of graphene, due to small values of $|\alpha|$ in graphene, TE surface waves are very weakly bounded ($|q_s|\sim1$)
and therefore they virtually do not couple to the point emitter. This means that, in practice, the contribution from TE pole can be neglected in graphene \cite{NikitinPRB11}. Nevertheless, taking into account a wide range of possible 2D sheets, in our analysis we retain the contribution from both poles.

In order to proceed with the series expansion, the singular terms in the integrands must be separated. The separation for the dyadics $\hat{\Phi}_\tau^{(n)}(w)$ into a pole and a smooth part $\hat{\Phi}_{S\tau}^{(n)}(w)$ is as follows:
\begin{equation}\label{a11}
\begin{split}
&\hat{\Phi}_\tau^{(n)}(w) = \frac{\hat{Q}_\tau^{(n)}}{w-w_\tau} + \hat{\Phi}_{S\tau}^{(n)}(w), \\
&\hat{\Phi}_{S\tau}^{(n)}(w) = \frac{\hat{\Phi}_\tau^{(n)}(w)(w-w_\tau)-\hat{Q}_\tau^{(n)}}{w-w_\tau},
\end{split}
\end{equation}
where $\hat{Q}_\tau^{(n)}$ are dyadics with the elements corresponding to the residues of $\hat{\Phi}_\tau^{(n)}(w)$, which after some algebra can be computed from Eq.~\eqref{a8} as:
\begin{equation}\label{a12}
\begin{split}
\hat{Q}_p^{(n)} = \frac{1}{\alpha^2 q_p}\hat{A}_p^{(n)}(q_p),\q
\hat{Q}_s^{(n)} = \dfrac{\alpha}{q_s}\hat{A}_s^{(n)}(q_s).
\end{split}
\end{equation}
Let us explicitly write out the expressions for the residue dyadics
\begin{equation}\label{gr21}
\begin{split}
&\hat{Q}_p^{(1)} = \frac{\sqrt{q_p}}{\alpha^2}
\begin{pmatrix}
\frac{-1}{\alpha q_p} & 0 & 1\\
0 & 0 & 0\\
1 & 0 & -\alpha q_p\\
\end{pmatrix}, \\
&\hat{Q}_s^{(1)} =\frac{-\alpha}{\sqrt{q_s}}
\begin{pmatrix}
0 & 0 & 0\\
0 & 1 & 0\\
0 & 0 & 0\\
\end{pmatrix},\\
&\hat{Q}_p^{(2)} = \frac{1}{8\alpha^2\sqrt{q_p}}
\begin{pmatrix}
\frac{-7}{\alpha q_p} & 0 & 3\\
0 & \frac{-8}{\alpha q_p} & 0\\
3 & 0 & \alpha\\
\end{pmatrix}, \\
&\hat{Q}_s^{(2)} =\frac{\alpha}{q_s^{\frac{3}{2}}}
\begin{pmatrix}
1 & 0 & 0\\
0 & \frac{-7}{8} & 0\\
0 & 0 & 0\\
\end{pmatrix}.
\end{split}
\end{equation}
Now we can deform the integration contour $C'$ in the $w$ complex plane, into the real $w$ axis. Then the singular terms $\frac{\hat{Q}_\tau^{(1,2)}}{w-w_\tau}$ in Eq.~\eqref{a11} for $\hat{\Phi}_\tau^{(1,2)}(w)$ that enter to the integral \eqref{a11} can be integrated analytically. The result for $\hat{g}_\tau(r,z)$ can be presented in the form of a sum
\begin{equation}\label{a13}
\begin{split}
\hat{g}_\tau(r,z) = i\pi\,e^{\rho(i-w_\tau^2)}\, \mathrm{erfc}(- iw_\tau\sqrt{\rho})\, \left(\hat{Q}_\tau^{(1)}+\frac{i}{r}\hat{Q}_\tau^{(2)}\right) \\
+ \hat{g}_{0\tau}(r,z),
\end{split}
\end{equation}
where the term with the complementary error function $\mathrm{erfc}(x)= (2/\sqrt{\pi}) \int_x^{\infty} e^{-t^2} dt $ is due the singularity. This function includes the contribution from the pole that is automatically taken into account when the pole is crossed by the transformation of the integration contour. The second term in Eq.~\eqref{a13}, $\hat{g}_{0\tau}(r,z)$ presents a nonsingular contribution
\begin{equation}\label{a13.1}
\begin{split}
\hat{g}_{0\tau}(r,z) = e^{i\rho}\int\limits_{-\infty}^{\infty} dw\, e^{-\rho w^2}\left[\hat{\Phi}_{S\tau}^{(1)}(w)+\frac{i}{r}\hat{\Phi}_{S\tau}^{(2)}(w)\right].
\end{split}
\end{equation}
The integral appearing in Eq.~\eqref{a13.1} is of the Gauss type, so the functions $\hat{\Phi}_{S\tau}^{(n)}(w)$ can be expanded in Tailor series close to $w=0$ and integration of every term is easily performed with the following result
\begin{equation}\label{a14}
\begin{split}
\hat{g}_{0\tau}(r,z) = e^{i\rho}\sum\limits_{m\in\mathrm{even}}\frac{1}{m!}\frac{\Gamma(\frac{1+m}{2})}{\rho^\frac{1+m}{2}}\\
\frac{d^m}{dw^m}\left[\hat{\Phi}_{S\tau}^{(1)}(w)+\frac{i}{r}\hat{\Phi}_{S\tau}^{(2)}(w)\right]_{w=0},
\end{split}
\end{equation}
where $\Gamma$ is Gamma function. Retaining in this expression the terms up to order $r^{-3/2}$ (which is enough for the most of the cases) yields
\begin{equation}\label{a15}
\begin{split}
\hat{g}_{0\tau}(r,z) \simeq e^{i\rho}\sqrt{\frac{\pi}{\rho}}\left[\hat{\Phi}_{S\tau}^{(1)}(0)+\frac{i}{r}\hat{\Phi}_{S\tau}^{(2)}(0)+\frac{1}{4\rho}\frac{d^2\hat{\Phi}^{(1)}_{S\tau}}{dw^2}|_{w=0}\right].
\end{split}
\end{equation}
Recalling that the saddle point $w=0$ corresponds to $\varphi=\theta$ and therefore to $q=\sin\theta$, we can explicitly write out the dyadics $\hat{\Phi}_{S\tau}^{(n)}(0)$
\begin{equation}\label{a16}
\begin{split}
\hat{\Phi}_{S\tau}^{(n)}(0)= \frac{\hat{Q}_\tau^{(n)}}{w_\tau} + \sqrt{2}e^{-i\frac{\pi}{4}}\frac{\hat{A}_\tau^{(n)}(\sin\theta)}{f_{\tau}(\sin\theta)}\cos\theta,
\end{split}
\end{equation}
where we have taken into account the identity $\frac{d\varphi}{dw} = \sqrt{2}e^{-i\frac{\pi}{4}}/\cos(\frac{\varphi-\theta}{2})$. The explicit expression for the third term in \eqref{a15} is more cumbersome for arbitrary $\theta$, therefore we give its formal expression, involving derivatives of previously defined functions
\begin{equation}\label{a17}
\begin{split}
&\frac{d^2\hat{\Phi}^{(1)}_{S\tau}}{dw^2}|_{w=0} = \frac{2\hat{Q}_\tau^{(1)}}{w_\tau^3} + \frac{d^2\hat{\Phi}^{(1)}_{\tau}}{dw^2}|_{w=0},\\
&\frac{d^2\hat{\Phi}^{(1)}_{\tau}}{dw^2}|_{w=0} = 2\sqrt{2}e^{-i\frac{3\pi}{4}}\frac{d^2}{d\varphi^2}\left[\frac{\cos\varphi}{\cos(\frac{\varphi-\theta}{2})}\frac{\hat{A}_\tau^{(1)}(\sin\varphi)}{f_{\tau}(\sin\varphi)}\right]_{\varphi=\theta}.
\end{split}
\end{equation}
The expressions \eqref{a13}, with $\hat{g}_{0\tau}(r,z)$ given by \eqref{a15}, present the analytical approximation of the DGF. Let us now analyze different terms in this expression and the applicability of the approximation.

\section{Analysis of the analytical solution}

In its general form, the analytical solution presents a non-trivial combination of the contribution from the pole and saddle point. The ``interaction'' between these contributions depends both upon the distance between the saddle point and the pole in the complex plane and upon the physical distance $r$ responsible for the oscillations of the integrand. The parameter that measures the interaction between the saddle point and the pole is called by Sommerfeld ``numerical distance''  $d_\tau$ and its square presents the argument of the complementary error function, $d_\tau^2 =- iw_\tau\sqrt{\rho}$.

In the $q$-plane, the saddle point corresponds to the condition of the extremum
of the exponential phase in \eqref{a5}, i.e. to $(q/q_z)_{min} = r/z$, or $q_{min} = \sin\theta$. This can be considered as an equation for the rays in ``Ray Optics'' (RO). If the contribution of the pole is neglected, then the same result can be derived following the standard stationary phase evaluation. In the far field, the leading RO contribution corresponds to the first term in \eqref{a15} with $\hat{\Phi}_{S\tau}^{(1)}(0)$ replaced by $\hat{\Phi}_{\tau}^{(1)}(0)$. It reads
\begin{equation}\label{a18}
\begin{split}
\hat{g}^{RO}_{\tau}(r,z) = e^{i\rho-i\frac{\pi}{4}}\sqrt{\frac{2\pi}{\rho}}\frac{\hat{A}_\tau^{(n)}(\sin\theta)}{f_{\tau}(\sin\theta)}\cos\theta.
\end{split}
\end{equation}
Returning to the DGF via Eq.~\eqref{a4}, we have explicitly [using Eqs.~\eqref{a5.2}, \eqref{a6} and the relation $r=\rho \sin\theta$]
\begin{equation}\label{a18.1}
\begin{split}
&\hat{G}^{RO}_p(\rho,\theta) = \frac{k_\omega e^{i\rho}\cos\theta}{4\pi \rho(1+\alpha\cos\theta)}
\begin{pmatrix}
\cos\theta & 0 & -\sin\theta\\
0 & 0 & 0\\
-\sin\theta & 0 & \frac{\sin^2\theta}{\cos\theta}\\
\end{pmatrix}, \\
&\hat{G}^{RO}_s(\rho,\theta)  =\frac{k_\omega e^{i\rho}\cos\theta}{4\pi \rho(\alpha+\cos\theta)}
\begin{pmatrix}
0 & 0 & 0\\
0 & 1 & 0\\
0 & 0 & 0\\
\end{pmatrix}.
\end{split}
\end{equation}

When the graphene sheet disappears,  $\alpha\rightarrow0$, we recover the spherical wave term ($\sim1/\rho$) of DGF corresponding to a dipole in FS. Notice that at $\theta=\pi/2$ (for the fields along $z=0$), the elements $rr$, $rz$, $zr$ in $\hat{G}^{RO}_{p}$ vanish independently upon whether the graphene sheet is present or not. In contrast, the element $\phi\phi$ in $\hat{G}^{RO}_{s}$ at $\theta=\pm\pi/2$ is very sensitive to the presence of graphene. It takes non-zero values for free space, $\alpha=0$ and vanishes for $\alpha\neq0$. This property of the $\phi\phi$ element is similar to the diffraction shadow effect in metals due to the presence of surface modes and formation of the Norton waves\cite{NikitinNJP09}. In case of metals, however, the diffraction shadow appears for the TM part of the dyadic, while in thin films this takes place for the TE part.

The fact that some elements of the dyadic $\hat{G}^{RO}$ vanish indicates that other terms in the general solution for $\hat{G}$ must be considered in order to provide the correct far-field representation of the DGF and electric fields. Let us consider in details the case of $\theta=\pi/2$.

\subsection{Asymptotic expansion of DGF in graphene plane, $z=0^-$~($\theta=\pi/2$)}

Here we present simplified expressions for $\hat{g}_{\tau}(r,z)$ at $z=0$. Recall that the general expression is given by the Eq.~\eqref{a13}, with $\hat{g}_{0\tau}$ given by Eq.~\eqref{a15}. The functions $\hat{\Phi}_{\tau}^{(n)}(0)$ at $\theta=\pi/2$ read
\begin{equation}\label{a19}
\begin{split}
&\hat{\Phi}_{s}^{(1)}(0)=\hat{\Phi}_{s}^{(2)}(0) = 0,\\
&\hat{\Phi}_{p}^{(1)}(0) = \sqrt{2}e^{-i\frac{\pi}{4}}\hat{z}\hat{z}, \q \hat{\Phi}_{p}^{(2)}(0) = -\frac{\sqrt{2}}{8}e^{-i\frac{\pi}{4}}\hat{z}\hat{z}.
\end{split}
\end{equation}
Introducing for a brevity of notations the following dyadic
\begin{equation}\label{a20}
\begin{split}
&\frac{d^2\hat{\Phi}^{(1)}_{s}}{dw^2}|_{w=0} \equiv \hat{M}_\tau \q \mathrm{for} \q \theta=\frac{\pi}{2},\\
\end{split}
\end{equation}
The derivatives \eqref{a17} simplify as
\begin{equation}\label{a20.1}
\begin{split}
&\hat{M}_p = 4\sqrt{2}e^{-i\frac{3\pi}{4}}
\begin{pmatrix}
1 & 0 & \alpha\\
0 & 0 & 0\\
\alpha & 0 & \alpha^2-\frac{9}{8}\\
\end{pmatrix},\\
&\hat{M}_s = \frac{4\sqrt{2}}{\alpha^2}e^{i\frac{\pi}{4}}
\begin{pmatrix}
0 & 0 & 0\\
0 & 1 & 0\\
0 & 0 & 0\\
\end{pmatrix}.
\end{split}
\end{equation}

Now we can explicitly write the full expression for the TE dyadic
\begin{equation}\label{a21}
\begin{split}
&\hat{G}_s(r,0) = \frac{k_\omega e^{i\frac{3\pi}{4}}}{8\pi}\sqrt{\frac{2\pi}{r}}\left(\hat{Q}_s^{(1)}+\frac{i}{r}\hat{Q}_s^{(2)}\right)e^{iq_sr} \mathrm{erfc}(- iw_s\sqrt{r}) \\
&+\frac{k_\omega e^{ir+i\frac{\pi}{4}}}{4\pi \sqrt{2}r}\left[\frac{1}{w_s}\left(\hat{Q}_s^{(1)}+\frac{i}{r}\hat{Q}_s^{(2)}\right)+\frac{1}{4r}\left(\frac{2\hat{Q}_s^{(1)}}{w_s^3} + \hat{M}_s\right)\right],
\end{split}
\end{equation}
and for the TM one
\begin{equation}\label{a22}
\begin{split}
&\hat{G}_p(r,0) = \frac{k_\omega e^{i\frac{3\pi}{4}}}{8\pi}\sqrt{\frac{2\pi}{r}}\left(\hat{Q}_p^{(1)}+\frac{i}{r}\hat{Q}_p^{(2)}\right)e^{iq_pr} \mathrm{erfc}(- iw_p\sqrt{r}) \\
&+\frac{k_\omega e^{ir+i\frac{\pi}{4}}}{4\pi \sqrt{2}r}\left[\frac{1}{w_p}\left(\hat{Q}_p^{(1)}+\frac{i}{r}\hat{Q}_p^{(2)}\right)+\sqrt{2}e^{-i\frac{\pi}{4}}\left(1+\frac{i}{8r}\right)\hat{z}\hat{z}\right.\\
&\left.+\frac{1}{4r}\left(\frac{2\hat{Q}_p^{(1)}}{w_p^3} + \hat{M}_p\right)\right].
\end{split}
\end{equation}
Recall that in Eqs.~\eqref{a21}-\eqref{a22} the residue dyadics $\hat{Q}_\tau^{(n)}$ are given by Eq.~\eqref{gr21}, and $\hat{M}_\tau$ have the form of Eq.~\eqref{a20.1}.
The locations of the poles in the complex $w$-plane at $\theta=\pi/2$ read
\begin{equation}\label{a10}
\begin{split}
w_{s,p}=e^{-i\frac{\pi}{4}}\sqrt{q_{s,p}-1}.
\end{split}
\end{equation}

We have checked that the Eqs.~\eqref{a21}, \eqref{a22} transform to the DGF of FS in the limit $\alpha\rightarrow0$. To perform this limit, one should carefully expand all the coefficients taking into account that for small $\alpha$ the poles become $w_s\simeq  \frac{\alpha}{\sqrt{2}}e^{i\frac{\pi}{4}}$, $w_p\simeq \frac{i}{\sqrt{\alpha}}$. In particular, for TM part of the DGF the expansion for large arguments of the complementary error function must be taken.

\subsection{Numerical check on the validity of the analytical approximation}

Let us present some illustrative examples that demonstrate the validity of the analytical approximation. For this we directly compare numerical and analytical calculations for two components of DGF in the most unfavorable situation, i.e. for $z=0$ ($\theta=\pi/2$), when the ``RO'' contribution disappears. We perform the precise (converged) numerical calculations according to Appendix A. The analytical approximation is given by the expressions \eqref{a21}, \eqref{a22}.

The conductivity of graphene is a function of frequency, $\nu=\omega/(2\pi)$, chemical potential $\mu$, temperature $T$ and scattering time $\tau$, see Appendix B. For the illustration, we consider the room temperature $T=300$K and $\mu=0.2$eV (48THz), which are typical experimental values.

\begin{figure}[tbh!]
\includegraphics[width=8.3cm]{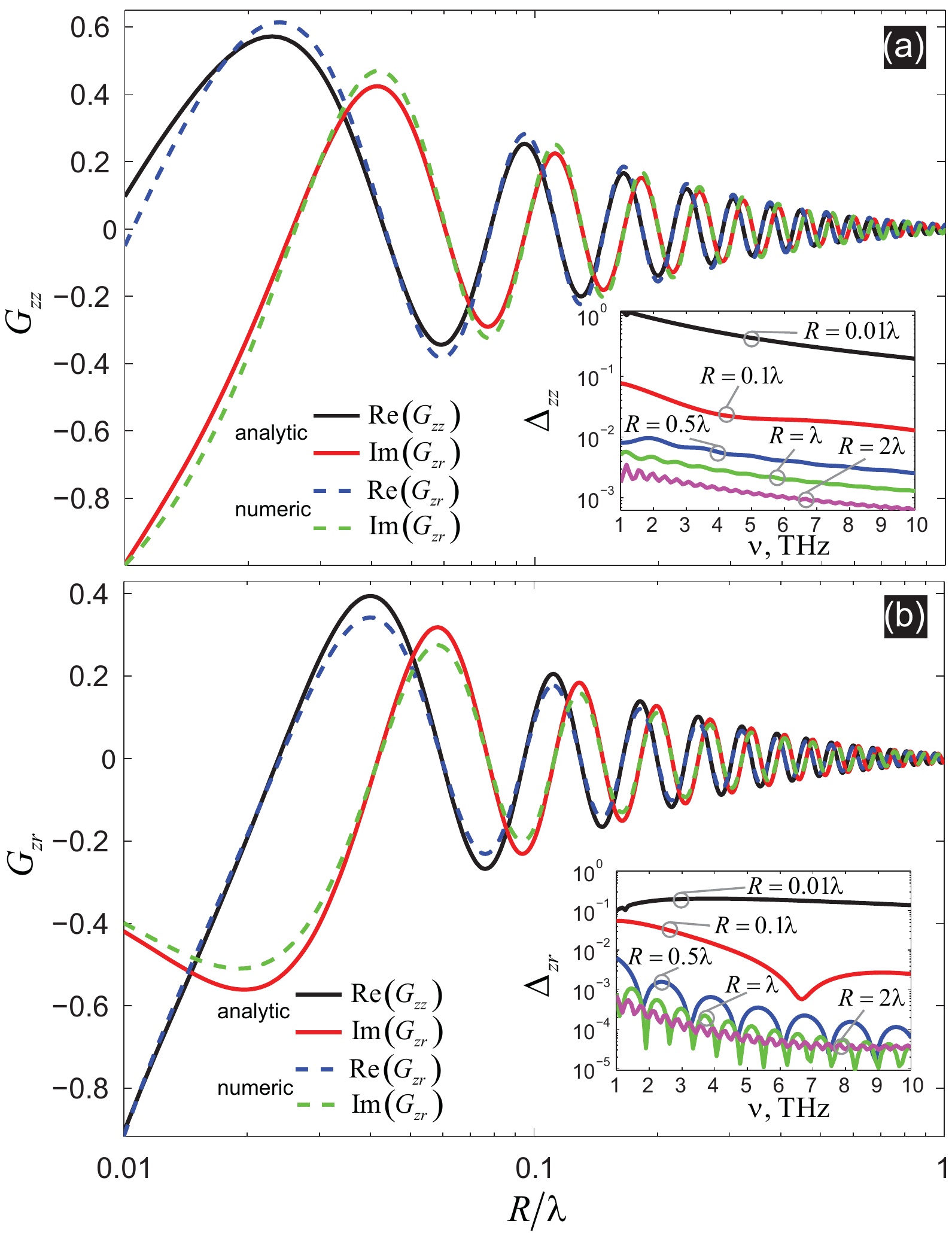}\\
\caption{Comparison between the numeric and analytic calculations of $zz$ and $zr$ DGF elements at $z=0$. The main figures show both real and imaginary parts of $G_{zz}$ and $G_{zr}$ as a function of distance. The values of these elements have been normalized to the maximal values of their modules in the shown range of distances.  The parameters for graphene in both panels are: $T=300$K, $\mu=0.2$ eV, $\tau=1$ps, $\nu=10$ THz. The insets show the dependencies of the relative error upon the frequency for different distances. In the insets  $\mu$ and $\tau$ are the same as in main figures.}\label{compar}
\end{figure}

The comparison is shown in Fig.~\ref{compar}. The range of the distances corresponds to the subwavelength region (outside of this region the the numerical and analytical curves are virtually undistinguishable).
In order to characterize the difference between numerical, $\hat{G}^{n}$, and analytical, $\hat{G}^{a}$, results, in the insets we have represented the relative error $\Delta_{\beta\beta'} = \left|\left(G_{\beta\beta'}^n-G_{\beta\beta'}^a\right)/G_{\beta\beta'}^n\right|$ in THz frequency range. According to the insets to Fig.~\ref{compar}, the error is a non-monotonous function of frequency. However, it has a decaying tendency with frequency increase. To understand such behavior, let us notice that for higher frequencies $|\alpha|$ decreases so that both the real and imaginary parts of $q_p$ increase. In the lower limit ($\nu=1$ THz) $\alpha\simeq0.11 + 0.69i$ so that $q_p\simeq 1.7 + 0.19i$,  while in the upper limit ($\nu=10$ THz) $\alpha\simeq0.0016 + 0.07i$ so that $q_p\simeq  14.34 + 0.34i$. Thus, the propagation length of the GSP, $L_{GSP} = \lambda/[2\pi\mathrm{Im}(q_p)]$, decreases due to increase of $\mathrm{Im}(q_p)$. This leads to a strong spacial decay of the terms in the solution (at the deep sub-wavelength distance), related to the GSP field components. Since the analytical solution recovers the FS DGF (up to $1/r^2$), the coincidence between the analytical and exact solutions improves for higher frequencies.

\section{Long distance limit. Surface modes and algebraically-decaying components}

In the region of parameters, where the argument of the complementary error function is a large number $|w_\tau|\sqrt{\rho}\gg1$ (the numerical distance is large, $|d^2_\tau|\gg1$), this function can be substituted by a few terms from its asymptotic expansion
\begin{equation}\label{a23}
\begin{split}
&\mathrm{erfc}(-iw_\tau\sqrt{\rho})= 2\Theta_-\left[\mathrm{Im}(w_\tau)\right] +\\
&\frac{e^{w_\tau^2 \rho}}{w_\tau \sqrt{\pi \rho}}\sum\limits_{n=0}^\infty\frac{(-1)^n}{(-i)^{2n+1}}\frac{(2n)!}{n!(2w_\tau )^{2n}\rho^n},
\end{split}
\end{equation}
where
\begin{equation}\label{a24}
\begin{split}
\Theta_-(x)=0, \q \mathrm{for} \q x\geq0,\\
\Theta_-(x)=1, \q \mathrm{for} \q x<0.
\end{split}
\end{equation}
The first term (which is independent upon $\rho$), appears when the transformation of the initial integration path to the steepest descent one results in crossing the pole.
Retaining the terms exact up to $r^{-3/2}$ in Eq.~\eqref{a23}, we arrive at
\begin{equation}\label{a25}
\begin{split}
&\mathrm{erfc}(-iw_\tau\sqrt{\rho})= 2\Theta_-\left[\mathrm{Im}(w_\tau)\right] +\\
&\frac{ie^{w_\tau^2 \rho}}{w_\tau \sqrt{\pi \rho}}\left(1 + \frac{1}{2w_\tau^2 \rho}\right) + O\left[(w_\tau\sqrt{\rho})^{-5}\right].
\end{split}
\end{equation}
Let us concentrate on the case of the in-plane fields, $\theta=\pi/2$.
Taking into account that $e^{w_\tau^2 r}e^{iq_\tau r} = e^{ir}$, the expressions \eqref{a21}, \eqref{a22} simplify to
\begin{equation}\label{a26}
\begin{split}
&\hat{G}_s(r,0) = \frac{k_\omega e^{i\frac{\pi}{4}}}{8\pi}\sqrt{\frac{2}{\pi r}}\left[\hat{\Pi}_s\,e^{iq_s r}+ e^{ir}\frac{\sqrt{\pi}}{4r\sqrt{r}}\hat{M}_s\right] + O\left(\frac{1}{r^3}\right),
\end{split}
\end{equation}
\begin{equation}\label{a27}
\begin{split}
&\hat{G}_p(r,0) = \frac{k_\omega e^{i\frac{\pi}{4}}}{8\pi}\sqrt{\frac{2}{\pi r}}\left[\hat{\Pi}_p\,e^{iq_p r} \right.\\
&+e^{ir-i\frac{\pi}{4}}\sqrt{\frac{2\pi}{r}}\left(1+\frac{i}{8r}\right)\hat{z}\hat{z}
+\left.e^{ir}\frac{\sqrt{\pi}}{4r\sqrt{r}}\hat{M}_p\right] + O\left(\frac{1}{r^3}\right),
\end{split}
\end{equation}
where the dyadic $\hat{\Pi}_\tau$ describes the surface mode and only contributes when the pole is located in the physically proper Riemann sheet
\begin{equation}\label{a28}
\begin{split}
\hat{\Pi}_\tau = 2\pi i\cdot\Theta_-\left[\mathrm{Im}(w_\tau)\right]\left(\hat{Q}_\tau^{(1)}+\frac{i}{r}\hat{Q}_\tau^{(2)}\right).
\end{split}
\end{equation}

We would like to notice that in the approximate expression for $\hat{g}_s(r,0)$ given by Eq.~\eqref{a26} we cannot recover the limit $\alpha\rightarrow0$ anymore.

\begin{figure*}[tbh!]
\includegraphics[width=16cm]{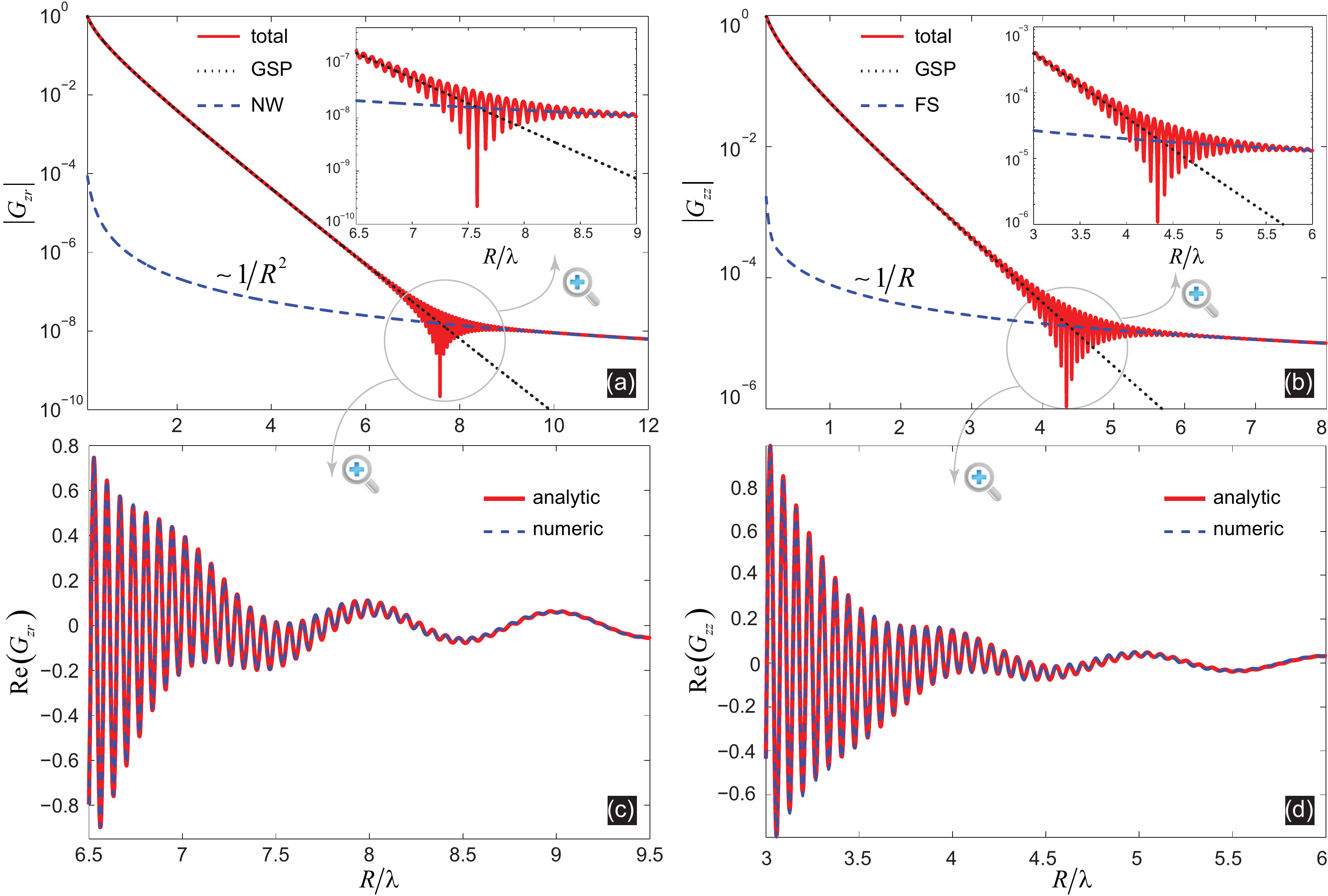}\\
\caption{(a,b) The absolute values of $G_{zr}$ and $G_{zz}$ as a function of the distance from the point source. The modulus of GSP, NW in (a) and GSP, NW in (b) terms are also rendered in the same panels.
The insets to (a,b) present zooms of the main panels in the region of a strong interference between GSP and algebraically-decaying components (FS and NW).
(c,d) The real part of $G_{zr}$ and $G_{zz}$ as a function of the distance from the point source in the region of interference between GSP and NW or GSP and FS. The analytical expressions based upon Eq.~\eqref{a27} are compared with the exact calculation. In both (a,b) and (c,d) all represented values are normalized to the maximal value of $|G_{zr}|$ ($|G_{zz}|$) in the shown intervals. The parameters of graphene are the same as in Fig.~\ref{compar}.}\label{NWGSP}
\end{figure*}

Expressions \eqref{a27}, \eqref{a28} present a sum of the surface mode term proportional to $\hat{\Pi}_\tau$ (GSP in case of $p$-polarization) and algebraically-decaying terms. The term describing the surface mode can be directly recovered from the angular representation Eq.~\eqref{a5}, considering only the residue of the pole. The algebraic components, in their turn can be derived from the same integral \eqref{a5}, considering the contribution from the branch-point $q_z=0$ (see Ref.~\cite{NikitinPRB11}). Asymptotic expressions \eqref{a27}, \eqref{a28} present thus \emph{independent} contributions from the pole and the branch cut.

The main physical reason of the validity of this approximation is that for sufficiently long distances only sharp peculiarities on the density of electromagnetic states (DES) contribute. DES is reflected by the integrand in Eq.~\eqref{a5}. For large $r$, the smooth region of (DES) is progressively canceled out in the integral, which is eventually dominated
by the strong (and rapid) contribution from the pole. The contribution of this pole gives the field of the surface mode. Due to losses, the density of states associated with the pole has a finite width, which causes the exponential decrease of the GSP amplitude with distance (characterized by the surface mode propagation length). Then, the contribution to the integral from either kink or square-root singularity ($\propto 1/q_z$) located at $q_z=0$ dominates. This takes place since this kind of features cannot be characterized by a typical width in $q$-space (these features are infinitely sharp in $q$-space), and they are not as strongly suppressed as the pole contribution when integrated with an oscillatory function. The contribution of the kink/square-root singularity yields the algebraic decay of the DGF components with respect to the distance.

A detailed physical description of the algebraically-decaying components of the fields from a point source can be found in Ref.~\cite{NikitinPRB11}. Let us recall here the physical meaning of all the algebraically-decaying terms in Eqs.~\eqref{a26}, \eqref{a28}. These terms appear in the dyadics $\hat{M}_\tau$, and the element $zz$ of $\hat{G}_p$, contains an additional contribution $\sim1/r$. The algebraically decaying components are composed of both FS terms and Norton waves (NW)~\cite{Norton36}.   The FS terms do not depend upon $\alpha$ and result from the contribution of the branch-cut singularity ($1/q_z$) that yields the dependency $\sim1/r$ and a kink that yields $\sim1/r^2$. Notice that while $zz$ component (of the TM part) contains both the term decaying as $\sim1/r$ and $1/r^2$, the FS part of the element $rr$ has only $1/r^2$ decay. All the rest of the $1/r^2$ terms that depend upon $\alpha$ correspond to the NW (compare with the case of metals, Ref.~\cite{NikitinNJP09,NikitinPRL10}). We would like to notice that as follows from Eqs.~\eqref{a26} (where $\alpha$ has been supposed to have a nonzero value) the element $\phi\phi$ contains only the NW. However, if we carefully perform the limit $\alpha\rightarrow0$ in the initial equation \eqref{a21}, recovering the FS DGF, the element $\phi\phi$ will contain a $\sim1/r$ term. This can be explained by the fact that in the DGF given by its angular representation \eqref{a5} has a square-root singularity for $\alpha=0$.

Since the main message of this paper is the analytical treatment of DGF for graphene, let us illustrate the validity of the asymptotic expressions \eqref{a27}, \eqref{a28} by comparing two elements ($zr$ and $zz$) of DGF with the numeric solution.

First, the competition between algebraically-decaying and GSP terms is shown in Fig.~\ref{NWGSP}~(a,b). At the beginning of the shown spacial window, both elements of DGF are dominated by the GSP terms. Then, in the region of $R\sim (7-9)\lambda$ for $G_{zr}$ and $R\sim (4-5)\lambda$ there is a crossover, where the exponentially decaying GSP is overcome by the algebraically decaying field. The algebraic decay for $G_{zr}$ corresponds only to NW ($\sim1/R^2$), since the FS contribution is zero for this element. In contrast, the asymptotic behaviour for $G_{zz}$ corresponds to FS with dominating $\sim1/R$ term. The NW term is also present in the element $zz$, but its contribution is much weaker than that of the FS component. In the region of the cross-over the field possesses a peculiar two-scaled oscillation behavior (corresponding to the wavelength of the GSP and vacuum wavelength). Notice that the amplitude of the field at the crossover is extremely small, so for this instance the analysis has mainly an academic value.

Second, in Fig.~\ref{NWGSP}~(c,d) a direct comparison of numeric and asymptotic results is performed in the interference region. As one can see, the asymptotic approximation perfectly captures all the details of the exact result. We have also checked the validity of our asymptotical expressions of all other DGF elements.

\section{Conclusion}

We have performed an analytical treatment of Dyadic Green's Function for 2D sheet. In particular, we have tested the analytical expressions on the case of graphene. We have retained all the necessary terms that provide high precision ($\sim 1\%$) down to distances of $1/10$ wavelengths and reasonable precision ($\sim 10\%$) down to $1/100$ wavelength.

For the limit of long distances (in units of plasmon wavelengths) we have presented simplified expressions with separated contribution from the pole (plasmonic field) and from the branch point (algebraically decaying field).
These expressions are relevant for future studies of the electromagnetic properties of subwavelengths objects placed on a graphene sheet.

\appendices

\section{Numerical computations of Sommerfeld integrals}

\begin{figure}[tbh!]
\includegraphics[width=8.3cm]{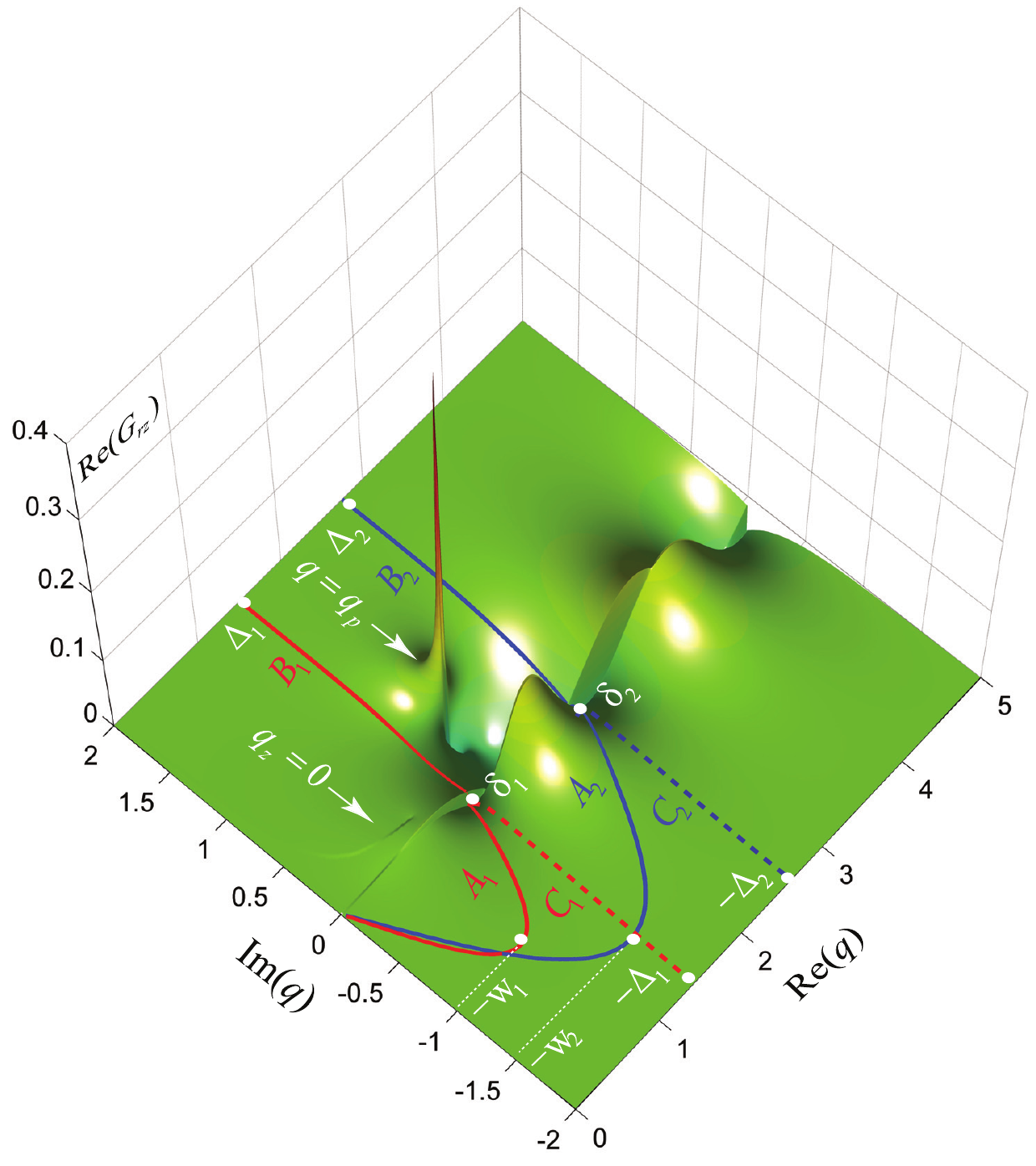}\\
\caption{The contours for the integrand corresponding to the DGF elements $G_{rz}=G_{zr}$ (the real part of the TM-term) for $z=0$, $r=0.5$. The integrand is normalized to the maximal value of its module. In the region $\mathrm{Im}(q)>0$ ($\mathrm{Im}(q)<0$) the Bessel function $J_1$ is replaced by the decaying Hankel function $H^{(1)}_1$ ($H^{(2)}_1$). As a result of this replacement, a discontinuity along $\mathrm{Im}(q)=0$ appears. The pole position $q_p=2 + 0.5i$ corresponds to a ``toy value'' of the normalized conductivity $\alpha\simeq0.168 + 0.516i$, chosen for better visualization. For the same reasons, in order to better illustrate the branch cut $\mathrm{Im}(q_z)=0$, the casuality has been exaggerated: $\sqrt{1-q^2}\rightarrow \sqrt{(1+i0)^2-q^2}\rightarrow \sqrt{(1+0.1i)^2-q^2}$.  The parameters of the contours: $\delta_1=1.5$, $\delta_2=2.75$; $w_1=1$, $w_2=1.5$; $\Delta_1=\Delta_2=2$.}\label{contour}
\end{figure}

Let us consider the integral of the following form
\begin{equation}\label{s11}
\begin{split}
I(r) = \int_0^\infty dq \mathcal{F}(q,r), \q  \mathcal{F}(q,r)=F(q)J_n(qr),
\end{split}
\end{equation}
where $J_n$ is the $n$th-order Bessel function and the function $F(q)$ remains finite for $\mathrm{Im}(q)\rightarrow \infty$. We suppose that the function $F(q)$ has a pole at $q=q_p$ and is dependent upon $q_z=\sqrt{1-q^2}$ so that it has branch cuts $\mathrm{Im}(q_z)=0$. The pole and the branch cut are not the only difficulties of the integral. In case of the integration along the real axis of the complex $q$-plane, the Bessel function has a strong oscillatory behavior for $q\gg1$ and the integration is very delicate. When $r$ increases, the convergence of the integral becomes worse. In order to stay away from the singularity and remain at the same Riemann sheet, the integration path can be deformed according to Cauchy theorem (supposing that we do not cross the pole)
\begin{equation}\label{s12}
\begin{split}
I(r) = \int_A dq F(q)J_n(qr) + \frac{1}{2}\int_B dq F(q)H^{(1)}_n(qr)\\
 + \frac{1}{2}\int_C dq F(q)H^{(2)}_n(qr).
\end{split}
\end{equation}
Here the contour $A$ passes below the real axis rounding the pole and the branch cut and then returns towards the real axis at the point $q=\delta$ with $\delta>\mathrm{Re}(q_p)$, moving into $\mathrm{Im}(q)\rightarrow \infty$ for $H^{(1)}_n$ term, and to $\mathrm{Im}(q)\rightarrow -\infty$ for $H^{(2)}_n$ (see Fig.~\ref{contour}, the paths marked by ``2''). The contours ``B'' and ``C'' are restricted by the limiting values $\mathrm{Im}(q)=\pm \Delta$.

When the pole is far away from the origin, $|q_p|\gg1$, or/and the distance parameter is large, $r\gg1$, it is convenient to bend the contours before the pole, i.e. choose $\delta<\mathrm{Re}(q_p)$ (see Fig.~\ref{contour}, the paths marked by ``1''). In this case the pole in the second integral of Eq.~\eqref{s12} must be taken into account. The contribution of the pole adds the residue term into the expression \eqref{s12}:
\begin{equation}\label{s13}
\begin{split}
I(r) = \int_Adq...  + \frac{1}{2}\int_Bdq... + \frac{1}{2}\int_Cdq...\\
 + \pi i \cdot\mathrm{Res}(F,q_p)H^{(1)}_n(q_pr).
\end{split}
\end{equation}

Each path ``A'', ``B'' and ``C'' can be parameterized $q_i=q_i(t)$ ($i=A,B,C$) so that the integration is reduced to the domain $[0,1]$:
\begin{equation}\label{s14}
I \simeq \sum_i\int\limits_{0}^1 dt\mathcal{F}[q_i(t)]\frac{dq_i}{dt},
\end{equation}
with
\begin{eqnarray}\label{s15}
&&q_A(t) = \delta \cdot t - i w\sin\left(\pi t\right), \nonumber\\
&&q_{B}(t) = \delta + it\Delta, \nonumber\\
&&q_{C}(t) = \delta - it\Delta,
\end{eqnarray}
where the parameters $\delta$, $w$ and $\Delta$ are chosen so that the best convergency of the integrals is provided.
In this paper, we have performed the integration over $t$ following Simpson's rule.

\section{Graphene's conductivity model}
The conductivity of graphene computed within the random phase approximation \cite{Wunsch06,Hwang07,Falkovsky08} can be written through the chemical potential $\mu$, the temperature $T$, and the scattering energy $\mathcal{E}_s$ as follows
\begin{equation}
\sigma = \sigma_{intra} + \sigma_{inter}
\end{equation}
where the intraband and interband contributions are:
\begin{equation}\label{s2}
\begin{split}
&\sigma_{intra} = \frac{2ie^2t}{\hbar\pi(\Omega+i\gamma)}\ln\left[2\cosh\left(\frac{1}{2t}\right)\right], \\
&\sigma_{inter} = \frac{e^2}{4\hbar}\left[\frac{1}{2} +\frac{1}{\pi}\arctan\left(\frac{\Omega-2}{2t}\right) -\right.\\
&\left. \frac{i}{2\pi}\ln\frac{(\Omega+2)^2}{(\Omega-2)^2+(2t)^2}\right].
\end{split}
\end{equation}
In this expressions $\Omega = \hbar\omega/\mu$, $\gamma=\mathcal{E}_s/\mu$ and $t = T/\mu $, with $T$ expressed in units of energy. The scattering energy is related to the relaxation time $\tau$ as $\tau=\mathcal{E}_s/\hbar$.


\section*{Acknowledgment}

The authors are grateful to Institute for Biocomputation and Physics of Complex Systems (BIFI) of Zaragoza for computational resources. We acknowledge support from the Spanish MECD under Contract No. MAT2011-28581-C02-02 and Consolider Project ``Nanolight.es''. FJGV acknowledges financial support by the European Research Council, grant 290981 (PLASMONANOQUANTA).

\ifCLASSOPTIONcaptionsoff
  \newpage
\fi


%
%
%

\end{document}